\documentstyle[prl,aps,psfig,multicol]{revtex}

\newcommand{\be}{\begin{equation}}
\newcommand{\ee}{\end{equation}}
\newcommand{\bea}{\begin{eqnarray}}
\newcommand{\eea}{\end{eqnarray}}
\newcommand{\bl}{\beta \lambda}
\newcommand{\w}{\omega}
\newcommand{\es}{\epsilon_s}
\newcommand{\eu}{\epsilon_u}

\begin{document}

 \title{The Adsorption and Collapse Transitions in a Linear Polymer Chain
near an Attractive Wall}
 \author{R. Rajesh$^1$, Deepak Dhar$^1$, Debaprasad Giri$^2$, Sanjay
Kumar$^3$ and Yashwant Singh$^3$}
 \address{
 {\small $^1$ Department of Theoretical Physics, Tata Institute of
Fundamental Research, Homi Bhabha Road, Mumbai-400005, India}\\
 {\small $^2$ Centre for Theoretical Studies, I. I. T., Kharaghpur-721302,
W. B., India}\\
 {\small $^3$ Department of Physics, Banaras Hindu University, Varanasi
221005, U. P., India}}
 \date{\today}
 \maketitle
 \widetext \begin{abstract}
 We deduce the qualitative phase diagram of a
long flexible neutral polymer chain immersed in a poor solvent near an
attracting surface using phenomenological arguments. The actual positions of
the phase boundaries are estimated numerically from series expansion up to
19 sites of a self-attracting self avoiding walk in three dimensions. In
two dimensions, we calculate analytically phase boundaries in some cases
for a partially directed model. Both the numerical as well as analytical
results corroborate the proposed qualitative phase diagram.

\noindent PACS number(s) : 64.60.-i,68.35.Rh,05.50.+q
 \end{abstract}

\begin{multicols}{2}

\section{Introduction \label{sec1}}

The behaviour of flexible polymers in solution at large length scales is
independent of the chemical nature of the polymer and the solvent, and
these universal scaling properties are well understood in terms of the
renormalisation group approach \cite{gennes,CJ}. The polymer chain is
known to undergo a transition from a random-coil phase to a globular phase
as the temperature, or the pH of the solution is varied.  The model of a
self-avoiding walk on a lattice with nearest-neighbour attractive or
repulsive interactions provides a simple model to understand the collapse
transition in polymers \cite{vanderzande}.

When the chain interacts with an impenetrable surface its conformational
properties are strongly modified \cite{eisen,BL}. There is a competition
between the lowering of internal energy near an attractive wall, and the
loss of entropy due to constraints imposed by the impenetrable surface.  
For a strongly attractive surface, the polymer sticks to the surface, and
for weak attraction, the preferred state is away from the surface. Thus,
there is a transition from the state when the chain is mostly attached to
the surface to a detached state, when the temperature is increased.
This behaviour finds applications in lubrication, adhesion, surface
protection, etc \cite{napper}.

If there is self-attraction as well as attraction to the surface, there is
the possibility of a collapse transition in the desorbed polymer, or 
in the polymer adsorbed on the surface.  In addition,
there is a surface-attached globular (SAG) phase, in which the polymeric
globule gets attached to the attractive surface \cite{SKG1}.  In the
thermodynamic limit, the SAG phase has the same free energy per monomer as
the bulk globular phase, and the transition between them is a surface
transition. In earlier papers \cite{SKG1,SKG2}, we had discussed the phase
diagram in this case, and investigated the phase diagram in a lattice
model using extrapolation of exact series expansions. This scheme has been
found to give satisfactory results as it can take into account the
corrections to scaling. To achieve the same accuracy by the Monte Carlo
method, a chain of about two orders of magnitude longer than in the exact
enumeration method has to be considered \cite{GH}.

In this paper, we show that the qualitative features of the phase diagram
in three dimensions can be determined by simple phenomenological arguments.
In the case of a partially directed polymer in two dimensions, we
determine the exact phase diagram of the SAG phase. In this case, the
polymer has different behaviour depending on whether it is near the wall
perpendicular to the preferred direction (SAG1) or the wall parallel to
the preferred direction (SAG2). We determine the phase boundaries of SAG1
and SAG2 phases by calculating their orientation dependent surface energy.
We also determine the transition between SAG1 and SAG2 phases when both
walls are present. We also summarise our results of analysis of exact
series expansion in three dimensions which we have extended by two more
terms.

The paper is organised as follows. Sec.~\ref{sec2} contains the definition
of the model and of the various phases.  In Sec.~\ref{sec3}, we briefly
review earlier work before providing arguments for the qualitative nature
of the phase diagram in two and three dimensions. The phase diagram
obtained is compared with numerical results from series
expansion in Sec.~\ref{sec4}. Sec.~\ref{sec5} contains the analytical
results obtained for the partially directed model.

\section{Model and Definitions } \label{sec2}

A simple lattice model for a linear polymer in a poor solvent is a self
avoiding walk (SAW) on a regular lattice with an attractive interaction
energy $\eu$ between pairs of sites of the walk which are unit distance
apart but not joined by an step of the walk. The adsorbing surface is
modelled by restricting the walk to lie in a upper half plane and by
associating an attractive energy $\es$ with each monomer [site of the
walk] lying on the surface. In the partially directed self-avoiding walk
(PDSAW) in two dimensions, there is an additional restriction that the
walk cannot take steps in the negative $x$-direction.

We will work with the reduced variables $\w=e^{\beta \es}$ and $u=e^{\beta
\eu}$, where $\beta$ is the inverse temperature. For the sake of easy
reference, we now define all the phases that we will encounter later on,
at one place. Consider a polymer consisting on $N$ monomers whose one end is
attached to a fixed site on the surface. 
If $\es$ and
$\eu$ are small in magnitude, the polymer exists in the swollen
random-coil phase, away from the surface. In this phase, the mean radius
of gyration varies as $N^\nu$ where $\nu$ takes the self avoiding walk
value [$\nu\approx 0.588$ in 3D and $\nu=3/4$ in 2D]. The number of
monomers in contact with the surface is of order one in this case. We
shall call this phase the desorbed extended ({\bf DE}) phase.  If $\eu$ is
large and $\es$ is small, the polymer exists away from the wall as a
compact ball of finite density. In this case, the radius of gyration of
the polymer varies as $N^{1/d}$ in $d$-dimensions. We shall call this
phase the desorbed collapsed ({\bf DC}) phase. If the surface attraction
$\es$ is sufficiently large, the polymer sticks close to the surface. In
this case, a finite fraction of monomers are on the surface, and the
extent of the polymer perpendicular to the surface is finite. Along the
surface the polymer roughly acts as a polymer chain in $(d-1)$-dimensions.
Depending on whether the attractive self-interaction is large or small,
the polymer is in a collapsed phase, with its transverse size varying as
$N^{1/(d-1)}$, or in the extended phase with the transverse size varying
as $N^{\nu^\prime}$, where $\nu^\prime$ is the self avoiding walk exponent
in $(d-1)$-dimensions. We shall call these phases the adsorbed collapsed
({\bf AC}) and the adsorbed extended ({\bf AE}) phases respectively. In
addition to these phases, the polymer may exist as a collapsed globule of
finite density which sticks to the surface. In this case, the size of the
polymer in the directions transverse and perpendicular to the surface
varies as $N^{1/d}$ and the number of monomers in contact with the surface
varies as $N^{(d-1)/d}$. We shall call this phase the surface adsorbed
globule ({\bf SAG}) phase. Note that in $2$-dimensions, the AC and the AE
phases cannot be distinguished from one another.

The polymer undergoes a transition from the extended to the collapsed
phase when the temperature is varied. At the transition temperature
between the DC and the DE phases, called the $\theta$-point, the critical
behaviour is described by a tricritical point of the $O(n)$ ($n
\rightarrow 0$) spin system. At the $\theta$-point, $R_b\sim
N^{\nu_\theta}$ with $\nu_\theta = 4/7$ for 2D \cite{DS} and $1/2$ for 3D
\cite{gennes}. The transition from AE to AC is described by $\nu_\theta$
corresponding to one lower dimension.

\section{Qualitative phase diagram} \label{sec3}

We first briefly review earlier results obtained for the model when both
the monomer-monomer interaction as well as the interaction with the wall
are attractive. One of the earlier papers on such systems was by Bouchaud
and Vannimenus \cite{BV}, in which the the phase diagram of a polymer
living on a Sierpinski gasket was analytically derived. The phase diagram
consisted of the AE, DE and the DC phases. In \cite{FOT}, the phase
diagram in $2$-dimensions was obtained approximately by series expansions
and it was found to be qualitatively similar to that for the gasket. In
\cite{SKG1}, the possibility of the existence of the SAG phase in
$2$-dimensions was pointed out based on analysis of series expansions.
Evidence for the existence of a surface transition from the SAG to DC
phase was also presented. A variant of the model, the PDSAW model in
$2$-dimensions, has been more amenable to analytical calculations. For
PDSAW, in $2$-dimensions, the exact calculation of the phase boundary
between the collapsed and the extended phases \cite{foster,igloi,FY} has
been numerically confirmed in \cite{VYG}. The phase diagram thus obtained
is qualitatively similar to the undirected $2$-dimensional model. In
\cite{MS}, the existence of the SAG phase in the PDSAW has been argued
for, based on series expansion analysis.

The model is less studied in $3$-dimensions.  Monte Carlo simulations
\cite{VW} and series expansion analysis \cite{SKG2} on the cubic lattice
showed the existence of four phases: AE, AC, DE and DC.  While \cite{VW}
claimed the existence of two multicritical points, the earlier preliminary
results \cite{SKG2} obtained from series expansion seemed to support one
multicritical point. More careful analysis of the series, reported later
in the paper, shows that there are indeed two multicritical points. The
question of whether an SAG phase exists in $3$-dimensions or not has not
been addressed so far. Also, the possibility of surface transitions among
the collapsed phases has not been explicitly dealt with. Thus, in spite of
many earlier studies, the qualitative behaviour of the system is not fully
established.

\begin{figure}
 \begin{center}
 \leavevmode
 \psfig{figure=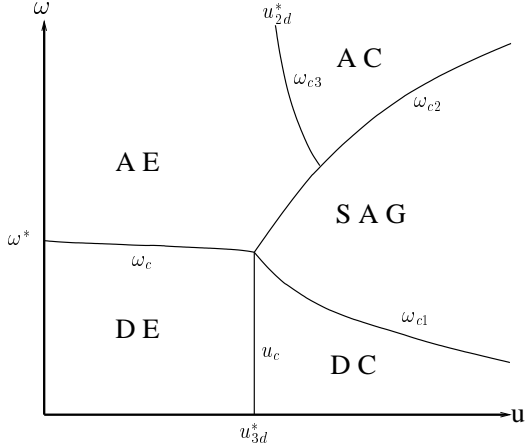,width=7cm,angle=0}
 \caption{The qualitative phase diagram in three dimensions.}
 \label{fig1}
 \end{center}
 \end{figure}

We now determine the qualitative nature of the phase diagram from
phenomenological considerations.  Consider the case when the interaction with
the wall is repulsive, i.e., $\w \leq 1$. The polymer will be in a
desorbed state because proximity to the surface results in increase of
internal energy as well as loss of entropy. As $u$ is increased from $1$
to $\infty$, the polymer undergoes a collapse transition from a desorbed
expanded state (DE) to a desorbed collapsed state (DC) at a critical value
$u^*_{3d}$ (see Fig.~\ref{fig1}). When the interaction with the wall is
attractive, there is a competition between lowering of free energy and
loss of entropy as the wall is approached. For $\es \protect \gtrsim 0$,
the loss of entropy is more dominant and hence the polymer remains
desorbed. Now, as $u$ is varied, the polymer undergoes a transition from
DE to DC at the same critical value $u^*_{3d}$.  Therefore, in the lower
part of the phase diagram, there is a vertical phase boundary $u=u_c$
separating the DE and the DC phases.

Consider now the case when the interaction between nearest neighbour
monomers is close to zero, i.e., $u \gtrsim 1$. Clearly, the polymer will
be in the DE phase. Now, as $\w$ is varied from $1$ to $\infty$ the
polymer undergoes a transition from DE to one in which it is adsorbed onto
the wall and extended, i.e., the AE phase.  Let this transition occur at a
critical curve $\w_{c}(u)$ that intersects the $\w$-axis at $\w^*$.

Now, consider possible phases of the system for large $\w$ and $u$. At
$T=0$, the polymer has the maximum possible density. It can then be
modelled as a Hamiltonian walk. The attractive energy per site is $-(d-1)
\eu$. In addition, for a finite polymer of $N$ monomers, there is a
surface energy cost which is easily seen to be $d \eu N^{(d-1)/d}$. Then
the free energy of the DC phase at $T=0$ is
 \be
 \mbox{E}_{\mbox{DC}}= -(d-1) \eu N+ d \eu N^{\frac{d-1}{d}}.
 \label{eq:1}
 \ee
 At $T=0$, for $\w_{c1}<\w<\w_{c2}$ (see Fig.~\ref{fig1}) 
the polymer exists in the SAG phase as a rectangular
parallelepiped of size $L_{\|}$ and $L_{\bot}$ in directions parallel and
perpendicular to the surface. Its bulk energy is the same as in the DC
phase and the surface energy is $(\eu-\es) L_{\|}^{d-1} + (d-1) \eu
N/L_{\|}$. Minimising the surface energy with respect to $L_{\|}$, we
obtain
 \be
 \mbox{E}_{\mbox{SAG}}= -(d-1) \eu N+ d \eu^{\frac{d-1}{d}} (\eu -
\es)^{\frac{1}{d}} N^{\frac{d-1}{d}}.
 \label{eq:2}
 \ee
 In the AC phase, we have $L_{\bot}=1$, $L_{\|}=N^{1/(d-1)}$ and the free
energy at $T=0$ is
 \be
 \mbox{E}_{\mbox{AC}} = -(d-2) \eu N -\es N+ (d-1) \eu
N^{\frac{d-2}{d-1}}.
 \label{eq:3}
 \ee
 From Eqs.~\ref{eq:1} and \ref{eq:2}, it follows that the surface energies
become equal when $\es=0$, i.e., $\w_{c1}\rightarrow 1$ as $u \rightarrow
\infty$. Also, the bulk free energy term for the AC phase has lower value 
than that of the SAG phase when $\es>\eu$. 
Thus, $\w_{c2}\rightarrow u$, as $u\rightarrow \infty$.

Consider now the case when $\w=\infty$. Then the polymer is adsorbed onto
the $(d-1)$-dimensional surface. In $d>2$, there is the additional
possibility of the AC phase. As $u$ is increased from $1$ to $\infty$, the
polymer will undergo a collapse transition, while still remaining
adsorbed, from AE to AC. This transition occurs at the critical value of
$(d-1)$-dimensional collapse $u^*_{d-1}$. Since the critical temperature
for collapse transition increases with dimension, $u^*_{d-1} > u^*_{d}$.
We will now argue that the curve $\w_{c3}$ originating from $(u^*_{2 d},
\infty)$ bends to the right when $\w$ is decreased to a finite but large
value. The partition function, when written as perturbation series in
$\w^{-1}$, is
 \be
 Z(u,\w)= Z_0(u) \w^N\! \left[1+\frac{N}{\w^2} \left(n_0+\frac{n_1}{u}+
 \frac{n_2}{u^2} \right)+\ldots\right]\!,
 \label{eq:4}
 \ee
 where $n_j$ is the fraction of bonds such that their end points have
exactly $j$ nearest neighbour monomers. Clearly, $n_0$ is larger in the AE
phase as compared to the AC phase, while $n_1$ and $n_2$ are smaller.
Using $n_0=1-n_1-n_2$ in Eq.~\ref{eq:4}, it follows that for large but
finite $\w$, the free energy is lower for the AE phase. Hence, the phase
boundary $\w_{c3}$ curves to the right.

The phase diagram for the $2$-dimensional problem is qualitatively the
same as that of the $3$-dimensional problem except that there is no AC
phase, and hence no $\w_{c3}$ phase boundary. We now argue that the phase
boundaries $u_c$, $\w_c$, $\w_{c1}$ and $\w_{c2}$ meet at one point. For
the sake of clarity, we will illustrate the arguments for the
$2$-dimensional problem. In the DC and the SAG phases, the polymer is a
compact two dimensional object with finite density.  We define
$\sigma(\theta)$ as the surface tension between the surface of this object
and the liquid, where $\theta$ is the angle the surface makes with the
horizontal. For a shape $r(\theta)$, the free energy is a sum of two
terms: the bulk term which depends on $u$ alone and a surface term, which
can be written as an integral over the angle dependent $\sigma(\theta)$.

Near the phase boundary $\w_{c2}$ separating the AE and the SAG phases,
the shape is highly anisotropic and $R_s \gg R_b$, where $R_s$ and $R_b$
are the extent of the polymer along and perpendicular to the surface.  
$R_s$ diverges as we approach the phase boundary from within the SAG
phase. Additional cost of creating two surfaces of orientation $\theta=0$
should be zero. Hence, along the phase boundary $\w_{c2}$, we have
 \be
 \sigma(0)+\sigma_w=0,
 \label{eq:5}
 \ee
 where $\sigma_w$ is the free energy cost per unit length when the polymer
is along the wall. Near the phase boundary $\w_{c1}$ separating the DC and
the SAG phases, the shape of SAG is such that the part in contact with the
wall has orientation $\theta=0$. Clearly, this configuration becomes
unfavourable in comparison to the DC phase when
 \be
 \sigma(0)=\sigma_w.
 \label{eq:6}
 \ee
 For the DE --- DC transition, clearly the surface tension must vanish at
the collapse point. Thus, along $u_c$ we have
 \be
 \sigma(0)=0.  \label{eq:7}
 \ee
 From Eqs.~\ref{eq:5}-\ref{eq:7}, it is clear that if any two of the phase
boundaries intersect at some point, then the third will also pass through
that point.

\begin{figure}
 \begin{center}
 \leavevmode
 \psfig{figure=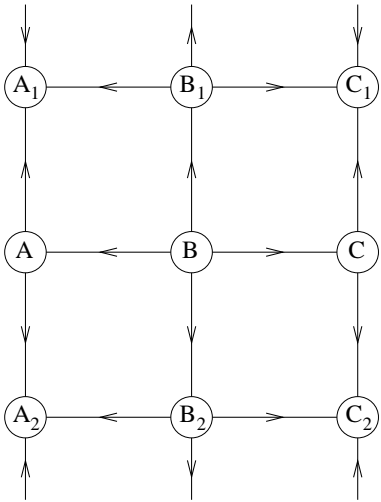,width=5cm,angle=0}
 \caption{The schematic flow diagram.}
 \label{fig2}
 \end{center}
 \end{figure}

It still remains to be shown that $\w_c$ will also pass through the same
point as the other phase boundaries. Let $u$ and $\w$ get transformed to
$u^{\prime}$ and $\w^{\prime}$ under a scale transformation as
 \bea
 u^{\prime}&=&f(u), \label{eq:8}\\
 \w^{\prime}&=&g(u,\w). \label{eq:9}
 \eea
 The function $f(u)$ is independent of the surface parameter $\w$ because
$u$ is a bulk parameter.  There will be three fixed points for
Eq.~\ref{eq:8} namely $u=0, u=u^*$ and $u=\infty$ where $u^*$ is the only
repulsive fixed point. Consider Eq.~\ref{eq:9} when $u$ is fixed at each
of its three fixed points. In the simplest scenario, for each value of
$u$, there will be three fixed points of Eq.~\ref{eq:9}, one corresponding
to no attraction, one to very strong attraction and the third a repulsive
fixed point. After fixing the flow directions, the final flow diagram
looks schematically as shown in Fig.~\ref{fig2}.  The fixed points $A_1,
A_2, C_1$ and $C_2$ correspond to the four phases. The fixed points $A,
B_1, C$ and $B_2$ correspond to the four critical phases corresponding to
the phase boundaries and the point $B$ corresponds to the repulsive
multicritical point.

\section{Series Expansion Results}\label{sec4}

We enumerated all SAWs up to a certain length on the cubic lattice in
which the first site of the walk lies at the origin and all other sites
are confined to the half plane $y\geq 0$.  Let $C_N (N_s, N_u)$ be the
number of SAWs of $N$ sites having $N_s$ monomers on $y=0$ and $N_u$
nearest neighbour monomer pairs. In \cite{SKG2}, we reported the
enumeration and analysis of the series $C_N (N_s, N_u)$ up to $N=17$ for
the cubic lattice.  We have now extended the series for $3$-dimensions by
two terms and reanalysed the data to obtain a better estimate of the phase
boundaries.

\begin{figure}
 \begin{center}
 \leavevmode
 \psfig{figure=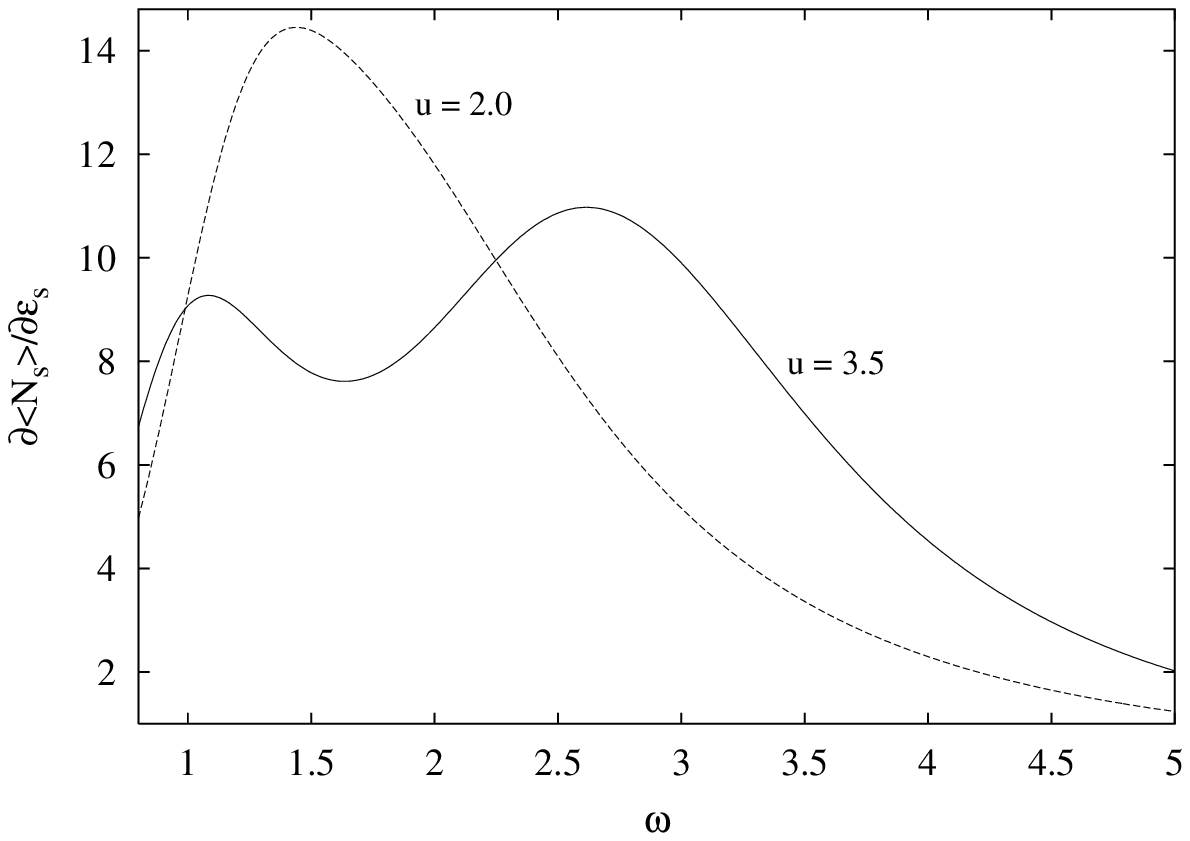,width=7cm,angle=0}
 \caption{The dependence of $\frac{\partial \langle N_s
\rangle}{\partial\es}$ on $\w$ is shown. For $u=2.0$, there is only one
peak corresponding to the DE to AE transition. For $u=3.5$, there are two
peaks corresponding to the DC to SAG to AC transition.}
 \label{fig3}
 \end{center}
 \end{figure}

For fixed $u$, we identify the position of the phase boundary separating
the desorbed phase from the adsorbed or attached phases as that value of
$\w$ at which $\frac{\partial \langle N_s \rangle}{\partial \es}$ is a
maximum. Fig.~\ref{fig3} shows the variation of $\frac{\partial \langle
N_s \rangle}{\partial \es}$ for two values of $u$ for $N=19$.

\begin{figure}
 \begin{center}
 \leavevmode
 \psfig{figure=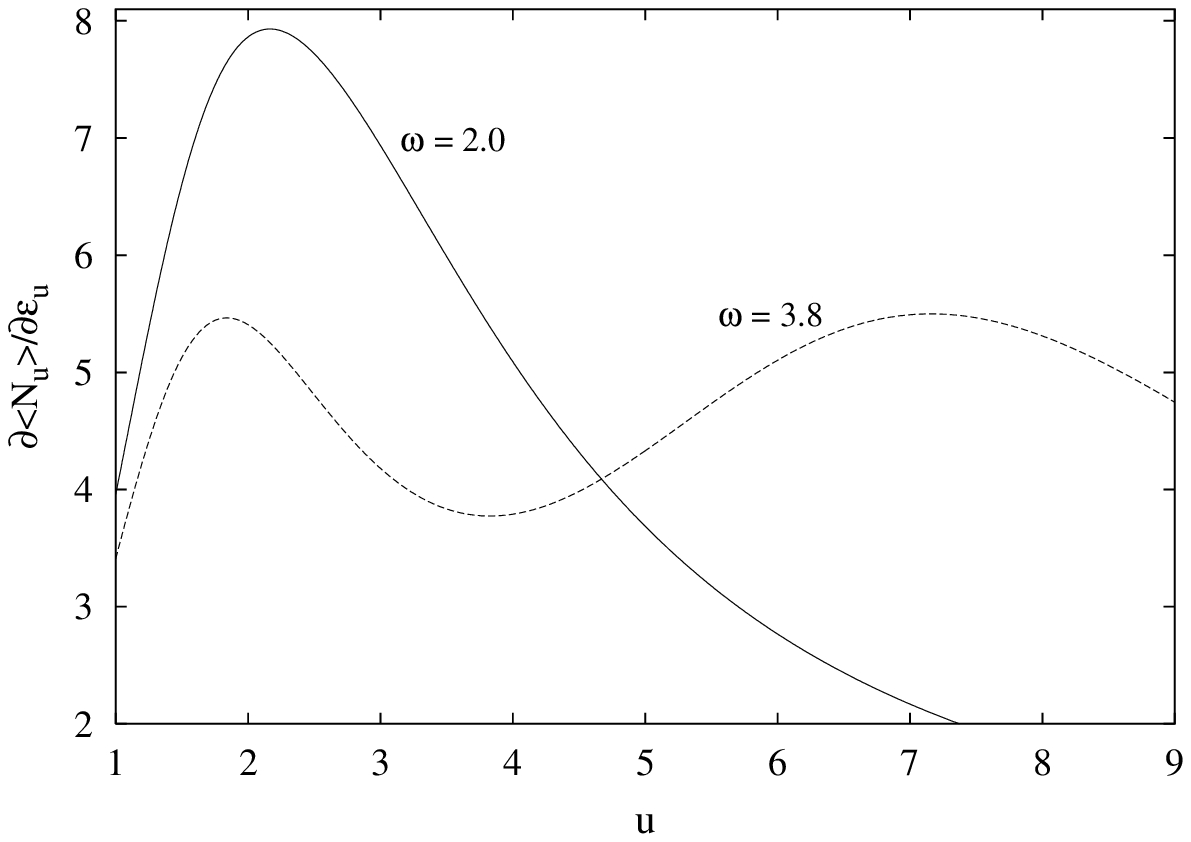,width=7cm,angle=0}
 \caption{The dependence of $\frac{\partial \langle N_u
\rangle}{\partial\eu}$ on $u$ is shown. For $\w=2.0$ there is only one
peak corresponding to the DE to DC transition. For $\w=3.8$, there are two
peaks corresponding to the AE to AC to SAG transition.}
 \label{fig4}
 \end{center}
 \end{figure}

For fixed $\w$, we identify the position of the phase boundary separating
the extended phase from the collapsed phase as that value of $u$ at which
$\frac{\partial \langle N_u \rangle}{\partial \eu}$ is a maximum.
Fig.~\ref{fig4} shows the variation of $\frac{\partial \langle N_u
\rangle}{\partial \eu}$ for two values of $\w$ for $N=19$.

The values of $u^*_{3d}$ and $\w^*$ obtained by this method is $2.00$ and
$1.49$ respectively. The previous results were $u^*_{3d}=1.76$ by series
expansion method \cite{SKG2} and $\w^*=1.45$ by Monte Carlo method
\cite{ma} and $\w^*=1.5$ by series expansion method \cite{TI}. It is
possible to obtain better estimates of $u^*_{3d}$ as well as the phase
boundaries by extrapolating for large $N$. Let
 \be
 Z_N(\omega, u) = \sum_{N_s, N_u} C_N(N_s, N_u) \omega^{N_s} u^{N_u},
 \label{eq:partition}
 \ee
 be the partition function. Then, the reduced free energy per monomer can
be written as
 \be
 G(\omega, u) = \lim_{N\rightarrow\infty} \frac{1}{N} \log Z_N (\omega,
u).
 \label{eq:freeenergy}
 \ee
 We refer to \cite{SKG1,SKG2} for details of the methods used for
extrapolating to large $N$ in Eq.~\ref{eq:freeenergy}. The phase
boundaries are then found from the maxima of $\frac{\partial^2
G(\omega,u)} {\partial \es^2}$ ($=\frac{\partial \langle N_s
\rangle}{\partial \es}$) and $\frac{\partial^2 G(\omega,u)}{ \partial
\eu^2}$ ($= \frac{\partial \langle N_u \rangle}{\partial \eu}$).

\begin{figure}
 \begin{center}
 \leavevmode
 \psfig{figure=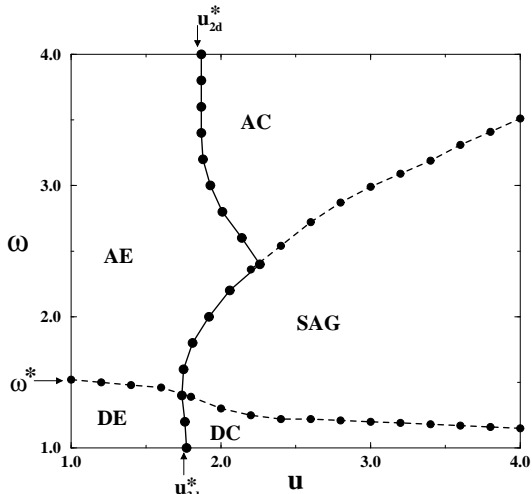,width=7cm,angle=0}
 \caption{The phase diagram for a polymer in 3-dimensions obtained from
series analysis is shown.The dashed and the full lines are the locations
of maxima of $\frac{\partial^2 G(\omega,u)}{\partial \es^2}$ and
$\frac{\partial^2 G(\omega,u)}{ \partial \eu^2}$ respectively.}
 \label{fig5}
 \end{center}
 \end{figure}

The phase diagram thus obtained is shown in Fig.~\ref{fig5}. We obtain
$u^*_{3d}=1.76$ and $\omega^*=1.48$ which accords fairly well with the
previously obtained results. The phase diagram obtained from series
analysis agrees qualitatively with the phase diagram proposed in
Sec.~\ref{sec3}.

\section{Analytic calculation for the two dimensional directed polymer}
\label{sec5}

In this section, we determine the phase boundary separating the SAG phase
from the DC and the AE phases in the PDSAW model. We do so by calculating
the macroscopic shape of the collapsed phases at low temperatures. At zero
temperature, it is easy to see that the configurational energy of the
polymer is minimised if it assumes a square shape of size $\sqrt{N}$ by
$\sqrt{N}$. For small nonzero temperatures, the polymer assumes a shape
which is slightly perturbed from this zero temperature square shape.  We
will derive an effective surface energy for these fluctuations in
Sec.~\ref{sec5_sub1}. Using these results, we determine the shape of SAG1
and SAG2 in Sec.~\ref{sec5_sub2}. In Sec.~\ref{sec5_sub3}, we calculate
the phase boundary between the various phases.

\subsection{Effective surface energy \label{sec5_sub1}}

\begin{figure}
 \begin{center}
 \leavevmode
 \psfig{figure=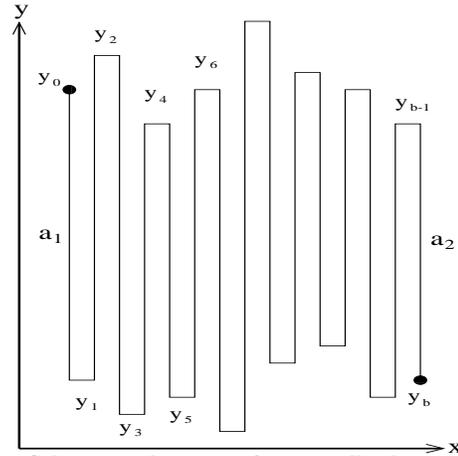,width=6cm,height=6cm,angle=0}
 \caption{Schematic diagram of a partially directed polymer for $T
\protect \gtrsim 0$}
 \label{fig6}
 \end{center}
 \end{figure}

For the directed polymer in the collapsed or the SAG phases, the density
in the bulk is exactly one and the configuration is ``frozen''. Only the
position of the boundary can change, as there is some fluctuation of
height allowed at the boundary. Thus $f_{\mbox{bulk}}(\mbox{SAG})
=f_{\mbox{bulk}}(\mbox{DC})=-\eu$, independent of $\w$. Consider a polymer
shape as shown in Fig.~\ref{fig6}.
 \noindent
 The energy of the configuration is
 \be
 E = -\eu N + \frac{\eu}{2}(a_1 + a_2+ 2 b) +\frac{\eu}{2}
\sum_{j=0}^{b-2} | y_{j+2}-y_{j} |.  \label{eq:12}
 \ee
 By a redefinition of $E$, we drop the bulk term proportional to $N$.  
The shape of the polymer is determined by the rest of the terms which are
all proportional to $\sqrt{N}$. We replace the terms under the summation
by an integral over an effective orientation dependent surface energy
$f(\theta)$, where $\theta$ is the angle the surface makes with the
horizontal. In this case, it is straightforward to calculate $f(\theta)$.
Consider all possible walks with an average slope $\tan(\theta)=y/x$.
Then, the sum over all weighted paths is
 \be
 e^{-\beta x \sec(\theta) f(\theta)}
 = \sum_{y_1,\ldots,y_x} \delta\left(\sum_{i=1}^{x} y_i -y\right)
\prod_{i=1}^x p^{ | y_i |},
 \label{eq:13}
 \ee
 where $p= e^{\frac{-\beta\eu}{2}}$ and $\delta$ is the usual Kronecker
delta function. Taking Laplace transform with respect to $y$, we obtain
independent summations over $y_i$. These are easily done giving
 \be
 f(\theta)=\frac{1}{\beta}\left[ \sin\theta
\log(z_0)+\frac{\cos\theta}{2}\log\frac{(z_0-p) (1-p z_0)} {z_0(1-p^2)}
\right],
 \label{eq:14}
 \ee
 where
 \be
 z_0=\frac{(1+p^2)\tan\theta + \sqrt{(1-p^2)^2 \tan^2 \theta+p^2}} {p(1+2
\tan\theta)}.
 \label{eq:15}
 \ee

We also need to calculate the energy cost $\sigma_w$ of adsorbing onto the
wall unit length of the polymer. For SAG1, it is trivially equal to
$\sigma_{w1}=-\es$. We calculate $\sigma_{w2}$ for SAG2 by the transfer
matrix method.  If $\psi_i$ denotes the $y$-coordinate of the lowest
portion of the polymer at site $i$, then the weight of obtaining
$\psi_{i+2}$ from $\psi_i$ is $\langle \psi_i | T |\psi_{i+2}\rangle =
\left[1+(\w^2-1)\delta_{\psi_{i+2}, 0}\right] u^{- | \psi_i-\psi_{i+2}
|/2}$. By trying out an ansatz $\psi_l=\alpha^l+\delta_{l,0} \psi_0$ for
the eigenfunction, it is not difficult to verify that the largest
eigenvalue of the transfer matrix T is
 \be
 \Lambda=\frac{\w^2(\w^2-1) (u-1)}{\w^2 (u-1)-u}.
 \label{eq:16}
 \ee
 Then, $\sigma_{w2}=-\log(\Lambda)/(2\beta)$. Clearly, as $u\rightarrow
\infty$, $\sigma_{w2}$ has the correct limit $-\es$.

\subsection{Calculation of the macroscopic shape \label{sec5_sub2}}

In this subsection we describe the shape determined by minimising the
surface energy of the collapsed phases.  Given the expression for the
temperature and orientation dependent $f(\theta)$, and also the value of
surface energy of polymer attached to the wall, it is straightforward to
determine the globular shape which minimises the surface energy given a
fixed volume. This is the classical Wulff construction. The result is that
the macroscopic shape of the polymer is given by
 \be
 e^{2 \bl y}= c_2 e^{\bl x} (1-p c_1 e^{-\bl x})(c_1 e^{-\bl x}-p),
 \label{eq:17}
 \ee
 where the two constants $c_1$ and $c_2$ are fixed by the two boundary
conditions. The Lagrange multiplier $\lambda$ is determined by the
constraint that the total area under the curves is $N$.  The constants
$c_1$ and $c_2$ are now varied to obtain the shape with the lowest surface
energy.

\begin{figure}
 \begin{center}
 \leavevmode
 \psfig{figure=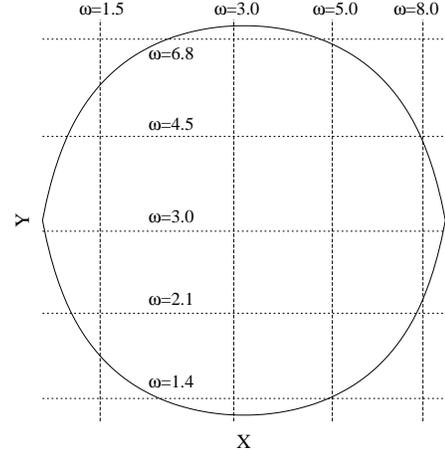,width=6.0cm,angle=0}
 \caption{The shape of SAG polymer is shown for different values of $\w$
when $u$ is kept fixed at $10.0$. The position of the wall is denoted by a
dotted line (vertical for SAG1 and horizontal for SAG2). The shape of SAG1
corresponds to the part of the curve from the wall to the right, while the
shape of SAG2 corresponds to part of the curve above the wall.}
 \label{fig7}
 \end{center}
 \end{figure}

Fig.~\ref{fig7} shows the shape of the SAG polymer for different values of
$\w$.  All the shapes lie on top of each other if we scale the coordinates
as $X=\bl x$ and $Y=\bl y$.

\subsection{Phase Diagram \label{sec5_sub3}}

We calculate the phase diagram for the directed polymer from
Eqs.~\ref{eq:5}-\ref{eq:7}. These equations give most of the phase
boundaries except the transition between SAG1 and DC and the transition
between SAG1 and SAG2. The SAG1-DC transition is not given by
Eq.~\ref{eq:6} because the shape in contact with the surface does not have
orientation $\theta=0$. This anomaly arises due to the constraint of
directedness. The surface transition from SAG1 to SAG2 is one in which the
globule would have lower free energy if attached to the $x$- wall rather
than the $y$-wall.

\noindent{\bf Transition from DC to DE $(u_c)$: } The critical value $u_c$
is obtained from Eq.~\ref{eq:7}, i.e, $\sigma(0)=0$.  This is equivalent
to the $2 f(0)+\eu=0$. Substituting for $f(0)$, we obtain
 \be
 \frac{\sqrt{u}-1} {\sqrt{u}+1}=\frac{1}{u},
 \label{eq:18}
 \ee
 which has the solution
 \be
 u_c = 3.38298\ldots .
 \label{eq:19}
 \ee
 Note that this result matches exactly with the result for DC-DE
transition obtained by the transfer matrix method \cite{foster,igloi,FY}.

\noindent{\bf Transition from SAG1 to AE $(\w_{c2})$ : } This phase
boundary is determined by equating the perpendicular extent of the SAG1
phase above the wall, to zero. This condition gives rise to the phase
boundary
 \be
 \w_{c2} = \frac{1+u^2 + \sqrt{(1+u^2)^2-4 u^3}}{2 u} .
 \label{eq:20}
 \ee
 This solution has a natural boundary at $u=u_c$ at which value the
expression under the square root sign becomes equal to zero.

\noindent{\bf Transition from SAG1 to DC $(\w_{c1})$ : } The transition
from SAG1 to DC occurs when the energy cost of creating a globule sticking
to the wall becomes equal to the energy of a DC polymer. We omit the
algebra and state the final result
 \be
 \w_{c1}(u)=\frac{1+u^2-\sqrt{(1+u^2)^2-4 u^3}}{2 u}.
 \label{eq:21}
 \ee

Previous analytical studies on the PDSAW \cite{foster,igloi,FY} had
considered the case when the wall was only along the $x$-direction. The
results obtained above for SAG1 are for a wall along the $y$-direction.
While the numerical values for the phase boundary differ, the phase
diagrams are qualitatively similar.

\noindent{\bf Transition from SAG2 to AE $(\w_{c2})$ : } From
Eq.~\ref{eq:5}, the phase boundary $\w_{c2}$ is given by
$\sigma(0)+\sigma_w=0$. Substituting the values of the surface energies
and solving for $\w$, we obtain
 \be
 \w_{c2}^2=\frac{ \alpha + \sqrt{\alpha^2-4u^3 }} {2 (1+\sqrt{u})},
 \label{eq:22}
 \ee
 where $\alpha=1+\sqrt{u}-u^2+u^{5/2}$. The phase boundary $\w_{c2}$ has a
natural boundary at $u=u_c$, at which value the expression under the
square root sign becomes equal to zero. The result differs from the
transfer matrix result \cite{foster,igloi,FY},
$\w_{c2}=\frac{u+1}{2}+\frac{\sqrt{(u^2+1)^2-4 u^3}}{2(u-1)}$. However,
this discrepancy is solely due to the fact that we consider only wall,
while the transfer matrix approach required two parallel walls. This
corresponds to changing Eq.~\ref{eq:5} to $2\sigma_{w2}+\eu=0$.
 
\noindent{\bf Transition from SAG2 to DC $(\w_{c1})$ : } From
Eq.~\ref{eq:6}, this transition occurs when $\sigma(0)=\sigma_w$. The
resulting equation can be solved to obtain
 \be
 \w_{c1}^2= \frac{\sqrt{u}} {\sqrt{u}-1}.
 \label{eq:23}
 \ee
 
This covers all the transitions when we consider SAG1 and SAG2 separately.
But if we consider the scenario where the possibility of both SAG's are
allowed, then there is a surface transition from one to the other when $u$
and $\w$ are varied.

\noindent {\bf Transition from SAG1 to SAG2 : } This transition is
determined by equating the surface energies of SAG1 and SAG2. However, it
turns out that we cannot obtain a closed form expression for the phase
boundary. Instead, we solved for it numerically using MATHEMATICA.

\begin{figure}
 \begin{center}
 \leavevmode
 \psfig{figure=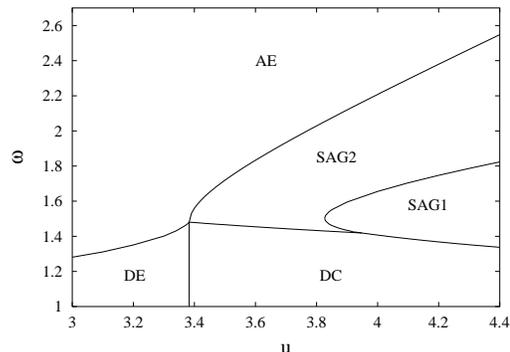,width=6.7cm,angle=0}
 \caption{The phase diagram for the $2$-dimensional PDSAW model. The DE-AE
phase boundary is schematic.}
 \label{fig8}
 \end{center}
 \end{figure}

In Fig.~\ref{fig8}, we plot the phase diagram when both SAG1 and SAG2 are
allowed to exist. Note that the phase diagram obtained is qualitatively
similar to the phase diagram proposed in Sec.~\ref{sec3}. The additional
transition between the SAG's is a consequence of the directed nature of
the PDSAW model.

The work at B. H. U.  was supported by the Department of
Science and Technology (India) through a project grant.

\end{multicols}


\begin{thebibliography}{1}

\bibitem{gennes} P. G. de Gennes, {\it Scaling concepts in polymer
Physics} (Cornell Univ. Press, Ithaca, 1979).

\bibitem{CJ} J. des Cloiseaux and G. Jannink, {\it Polymers in Solution}
(Clarendon, Oxford, 1980).

\bibitem{vanderzande} C. Vanderzande, {\it Lattice models of polymers}
(Cambridge University Press, UK, 1998).

\bibitem{eisen} E. Eisenriegler, {\it Polymers near surfaces} (World
Scientific, Singapore, 1993).

\bibitem{BL} K. D\'{e} Bell and T. Lookmann, Rev. Mod. Phys. {\bf 65}, 87
(1993).

\bibitem{napper} D. Napper, {\it Polymeric Stabilization of Colloidal
Dispersion} (Academic, New York, 1983).

\bibitem{SKG1} Y. Singh, D. Giri and S. Kumar, J. Phys. A {\bf 34}, L67
(2001).

\bibitem{SKG2} Y. Singh, S. Kumar and D. Giri, J. Phys. A {\bf 32}, L407
(1999).

\bibitem{GH} P. Grassberger and R. Hegger, Phys. Rev. E {\bf 51}, 2674
(1995); J. Phys. I (France) {\bf 5}, 597 (1995).

\bibitem{DS} D. Duplantier and H. Saleur, Phys. Rev. Lett. {\bf 59}, 539
(1987).

\bibitem{BV} E. Bouchaud and J. Vannimenus, J. Phys. France, {\bf 50},
2931 (1989).

\bibitem{FOT} D. P. Foster, E. Orlandini and M. C. Tesi, J. Phys. A {\bf
25}, L1211 (1992).

\bibitem{foster} D. P. Foster, J. Phys. A {\bf 23}, L1135 (1990).

\bibitem{igloi} F. Igl$\acute{o}$i, Phys. Rev. A {\bf 43}, 3194 (1991).

\bibitem{FY} D. P. Foster and J. M. Yeomans, Physica A {\bf 177}, 443
(1991).

\bibitem{VYG} A. R. Veal, J. M. Yeomans and G. Jug, J. Phys A {\bf 23},
L109 (1990).

\bibitem{MS} P. K. Mishra and Y. Singh, cond-mat/0009345.

\bibitem{VW} T. Vrobov$\acute{a}$ and S. G. Whittington, J. Phys. A {\bf
31}, 3989 (1998).

\bibitem{ma} L. Ma, K. M. Middlemiss, S. H. P. Bly and S. G. Whittington,
J. Chem. Soc. Faraday Trans. {\bf 74}, 721 (1978).

\bibitem{TI} T. Ishinabe, J. Chem. Phys {\bf 76}, 5589 (1982).

\end{thebibliography}
\end{document}